\documentclass[conference]{IEEEtran}
\IEEEoverridecommandlockouts
\usepackage{dblfloatfix}
\usepackage{setspace}
\usepackage{adjustbox}
\usepackage{cite}
\usepackage{amsmath,amssymb,amsfonts, pifont}
\usepackage{algorithmic}
\usepackage{graphicx}
\usepackage{textcomp}
\usepackage{xcolor}
\usepackage{listings}
\usepackage{color}
\usepackage{anyfontsize}

\usepackage{listings, xcolor}

\definecolor{verylightgray}{rgb}{.97,.97,.97}

\lstdefinelanguage{Solidity}{
	keywords=[1]{anonymous, assembly, assert, balance, break, call, callcode, case, catch, class, constant, continue, constructor, contract, debugger, default, delegatecall, delete, do, else, emit, event, experimental, export, external, false, finally, for, function, gas, if, implements, import, in, indexed, instanceof, interface, internal, is, length, library, log0, log1, log2, log3, log4, memory, modifier, new, payable, pragma, private, protected, public, pure, push, require, return, returns, revert, selfdestruct, send, solidity, storage, struct, suicide, super, switch, then, this, throw, transfer, true, try, typeof, using, value, view, while, with, addmod, ecrecover, keccak256, mulmod, ripemd160, sha256, sha3}, 
	keywordstyle=[1]\color{blue}\bfseries,
	keywords=[2]{address, bool, byte, bytes, bytes1, bytes2, bytes3, bytes4, bytes5, bytes6, bytes7, bytes8, bytes9, bytes10, bytes11, bytes12, bytes13, bytes14, bytes15, bytes16, bytes17, bytes18, bytes19, bytes20, bytes21, bytes22, bytes23, bytes24, bytes25, bytes26, bytes27, bytes28, bytes29, bytes30, bytes31, bytes32, enum, int, int8, int16, int24, int32, int40, int48, int56, int64, int72, int80, int88, int96, int104, int112, int120, int128, int136, int144, int152, int160, int168, int176, int184, int192, int200, int208, int216, int224, int232, int240, int248, int256, mapping, string, uint, uint8, uint16, uint24, uint32, uint40, uint48, uint56, uint64, uint72, uint80, uint88, uint96, uint104, uint112, uint120, uint128, uint136, uint144, uint152, uint160, uint168, uint176, uint184, uint192, uint200, uint208, uint216, uint224, uint232, uint240, uint248, uint256, var, void, ether, finney, szabo, wei, days, hours, minutes, seconds, weeks, years},	
	keywordstyle=[2]\color{teal}\bfseries,
	keywords=[3]{block, blockhash, coinbase, difficulty, gaslimit, number, timestamp, msg, data, gas, sender, sig, value, now, tx, gasprice, origin},	
	keywordstyle=[3]\color{violet}\bfseries,
	identifierstyle=\color{black},
	sensitive=false,
	comment=[l]{//},
	morecomment=[s]{/*}{*/},
	commentstyle=\color{gray}\ttfamily,
	stringstyle=\color{red}\ttfamily,
	morestring=[b]',
	morestring=[b]"
}
\usepackage{balance}

\begin{document}
\author{\IEEEauthorblockN{Noama Fatima Samreen, Manar H. Alalfi}
\IEEEauthorblockA{\textit{Department of Computer Science} \\
\textit{Ryerson University,
Toronto, ON, Canada} \\
{\{noama.samreen,manar.alalfi\}@ryerson.ca}}}
\title{{Reentrancy Vulnerability Identification in Ethereum Smart Contracts}
}
\vspace{-0.7 cm}

%
\maketitle
\vspace{-0.3 cm}
\begin{abstract}
Ethereum Smart contracts use blockchain to transfer values among peers on networks without central agency. These programs are deployed on decentralized applications running on top of the blockchain consensus protocol to enable people make agreements in a transparent and conflict free environment. The security vulnerabilities within those smart contracts are a potential threat to the applications and have caused huge financial losses to their users. In this paper, we present a framework that combines static and dynamic analysis to detect \textit{Reentrancy} vulnerabilities in Ethereum smart contracts. This framework generates an attacker contract based on the ABI specifications of smart contracts under test and analyzes the contract interaction to precisely report \textit{Reentrancy} vulnerability. We conducted a preliminary evaluation of our proposed framework on 5 modified smart contracts from Etherscan and our framework was able to detect the \textit{Reentrancy} vulnerability in all our modified contracts. 
Our framework analyzes smart contracts statically to identify potentially vulnerable functions and then uses dynamic analysis to precisely confirm \textit{Reentrancy} vulnerability, thus achieving increased performance and reduced false positives.
\end{abstract}
\section{Introduction}
Blockchain technology(BT) has been gaining popularity because of its wide range of potential applications. It was first applied as a cryptocurrency, called Bitcoin \cite{Bitcoin}, but has since been used in many other applications such as e-commerce, trade and commerce, production and manufacturing, banking, and gaming. BT uses a peer-to-peer (peers are known as miners in BT) framework which is a more decentralized approach to storing transaction and data registers. As there is no single point of failure or a third-party centralized control of transactions, BT has been standing out from other cryptocurrency technologies. It uses a chain of blocks in which each block is locked cryptographically using the hash of the previous block it is linked to, which creates an immutable database of all transactions stored as a digital ledger, and it cannot be changed without affecting all the blocks linked together in the chain\cite{surveyattacks}. 

Manipulations to the Blockchain is done using a proof of work (POW) system in which computers must solve a complex computational math problem to become eligible to add a block to the Blockchain. The theoretical aspect of using a cryptographically locked chain of blocks may seem to ensure data security and integrity from unauthorized access. However, on-going research on the security, integrity and authenticity of BT have shown that many applications have been exposed to an intrusion attack\cite{vulnerabilities}. 

One of the most destructive attacks in Solidity smart  contract is \textit{Reentrancy} attacks. A \textit{Reentrancy} attack occurs when the attacker drains funds from the target by recursively calling the target’s withdraw function. When the contract fails to update its state, a victim’s balance, prior to sending funds, the attacker can continuously call the withdraw function to drain the contract’s funds. A famous real-world \textit{Reentrancy} attack is the DAO attack which caused a loss of 60 million US dollars.

According to research by Liu et al.\cite{Reguard} detecting the existence of \textit{Reentrancy} vulnerabilities in Smart-Contracts faces two challenges:
\begin{itemize}
\item The implementation of Smart-Contracts varies widely and analysis to account for all possible scenarios may become infeasible.
\item The \textit{Reentrancy} vulnerability cannot be detected accurately because it lacks a definite pattern in the context of Smart-Contracts. Analyzing with a straightforward and simple pattern may result in false positives, while precise and rigorous patterns may fail to report the existence of a \textit{Reentrancy} vulnerability.
\end{itemize}
This paper presents a framework to address the \textit{Reentrancy} attack and provide solutions for the above challenges. This paper has the following contributions:
\begin{enumerate}
    \item A semi-automated framework with a combined static and dynamic analysis to better capture the various patterns of the \textit{Reentrancy} attack efficiently and accurately.
    \item A demonstration of the framework on the analysis of 5 smart contracts and comparison with one of the state of the art tools.
    \item A solidity smart contract TXL grammar which provides a fixable means to analyze various implementations and versions of smart contracts.
\end{enumerate}
\section{Background}
\vspace{-0.1 cm}
\subsection{Ethereum}
One of the Blockchain technology platforms is Ethereum\cite{Ethereum} which can implement algorithms expressed in a general-purpose programming language allowing developers to build a variety of applications, ranging from simple wallet applications to complex financial systems for the banking industry. These programs are known as Smart-Contracts which are written in a Turing-complete bytecode language, called EVM bytecode\cite{Ethereum}. The transactions sent to the Ethereum network by the users can create new contracts, invoke functions of a contract, and/or transfer ether to contracts.
\subsection{Smart Contract}
\vspace{-0.1 cm}
Ethereum uses Smart-Contracts\cite{Ethereum}, which are computer programs that directly controls the flow or transfer of digital assets.  Each function invocation in a Smart-Contract is executed by all miners in the Ethereum network and they receive execution fees paid by the users for it. Execution fees also protect against denial-of-service attacks, where an attacker tries to slow down the network by requesting time-consuming computations. This execution fee is defined as a product of “gas” and “gas-price”. 
Implementing a Smart-Contract use case can pose few security challenges, like public visibility of the complete source code of an application on a network, and validation and verification of the source code. Moreover, the immutability of Blockchain technology makes patching discovered vulnerabilities in already deployed Smart-Contracts impossible.
\begin{lstlisting}[caption={Solidity Contracts Sender and Reciever},label={lstNew}]
pragma solidity >=0.4.22 <0.6.0;
contract Sender {
    uint public amount;
    address payable public sender;
    address payable public reciever;
    constructor() public payable {
        sender = msg.sender;
        amount = msg.value;
    }
    function send(receiver) payable {
        receiver.call.value(value).gas(20317)();
    }
}
contract Receiver {
  uint public balance = 0;
  function () payable {
    balance += msg.value;
  }
}
\end{lstlisting}
\subsection{Solidity}
\vspace{-0.1 cm}
Smart contracts are typically written in a high-level Turing-complete programming language such as Solidity\cite{Solidity}, and then compiled to the Ethereum Virtual Machine (EVM) bytecode\cite{Ethereum}, a low-level stack-based language. For instance, Listing \ref{lstNew} shows a smart contract written in the Solidity programming language\cite{Solidity}.

In Listing \ref{lstNew}, the first line of the program declares the Solidity’s version. The program will be compatible with the corresponding EVM or any other higher version less than 0.6.0. It contains a constructor to create an instance of the contract, and functions.

 The \textit{msg.sender} is a built-in global variable representative of the address that is calling the function. The \textit{msg.value} is another built-in variable that tells us how much ether has been sent. The \textit{payable} keyword is what makes solidity truly unique. It allows a function to send and receive ether. 
 
 A function with no name followed by \textit{payable} keyword, \textit{function () payable{}}, is a fallback function in solidity that is triggered when a function call's identifier does not match any of the existing functions in a smart contract or if there was no data supplied at all. 

To transfer ether between the contracts, Solidity uses \textit{send()}, \textit{transfer()} and \textit{call()}. \textit{send()} transfers the ether and executes the fallback function of the contract that receives ether. The gas limit per execution is 2300 and an unsuccessful execution of \textit{send()} function does not throw an exception but the gas required for the execution is spent. Similarly, \textit{transfer()} is also used to transfer ether between the contract, but the gas limit can be redefined using the \textit{.gas()} modifier, and an unsuccessful \textit{transfer()} throws an exception. The gas limit in both these functions prevents the security risk involved in executing expensive state changing code in the fallback function of the contract receiving the ether. The one pitfall is when a contract sets a custom amount of gas using the \textit{.gas()} modifier.
The \textit{call()} function is comparatively more vulnerable as there is no gas limit associated with this function.

\section{Reentrancy Vulnerability}
\vspace{-0.1 cm}
According to Atzei et al. \cite{surveyattacks}, a \textit{Reentrancy} attack can drain a Smart-Contract of its ether and can aid an intrusion into the contract code. When an external function call another untrusted contract and an attacker gains control of this untrusted contract, they can make a recursive call back to the original function, unexpectedly repeating transactions that would have otherwise not run, and eventually consume all the gas.
\subsection{On A Single Function}
If a contract uses \textit{call}, \textit{send} or \textit{transfer} which may cause control flow to an external contract, with a fallback function, and then updates the state afterward then this causes the state of the contract to be incomplete when flow control is transferred (see Listing \ref{lst2}). Therefore, when this fallback function is triggered, the flow of control may not return as the called contract expects and the caller might do any number of unexpected things such as calling the function again, calling another function or even calling another contract.
\begin{lstlisting}[caption={Reentrancy vulnerability on a single function},label={lst2}]  
function transferBalance(address receiver, uint amount) 
 public {
  require(balances[msg.sender] >= amount);
  receiver.transfer(amount);
  /* flow control transferred before the sender's balance is updated and before an event is emitted. Potentially the start of trouble. */
  balances[receiver] -= amount;
  LogTransactions(msg.sender,receiver, amount); 
  }
 \end{lstlisting}
In the example given, one of the mitigation methods is to use send() which will limit, to an extent, the execution of expensive external code \cite{countermeasures}.
However, this can be completely avoided by updating \textit{balances} before transferring the control to another contract. That is, put the whole state in order before invoking another contract. If something goes wrong, revert everything to the original state (see Listing \ref{lst3}).
\begin{lstlisting}[caption={Reentrancy Vulnerability on a single function - Prevention},label={lst3}]  
function transferBalance(address receiver,uint amount)public{
    require(balances[msg.sender]>=amount);
    balances[msg.sender]-=amount;
    LogTransactions(msg.sender,receiver, amount);
    receiver.transfer(amount);
    //<==on fail, this will revert all the above.
  }
 \end{lstlisting}
\subsection{Cross-Function Reentrancy}
A similar attack can be done when two different functions or even contracts share the same state.
In this case (Listing \ref{lst4}), the attacker calls transfer() when their code is executed on the external call in \textit{withdraw}. Since their balance has not yet been set to 0, they are able to transfer the tokens even though they already received the withdrawal. This vulnerability was also used in the DAO attack\cite{vulnerabilities}. The same solution to update all the balances before transferring control to another function or contract will work even in this case.
\begin{lstlisting}[caption={Cross-function Reentrancy vulnerability},label={lst4}]
mapping (address => uint) private balance;
function transfer(address to, uint amount) {
    if (balance[msg.sender] >= amount){
      balance[to] += amount;
      balance[msg.sender] -= amount;
    }
  }
function withdraw() public {
   uint amount = balance[msg.sender];
   require(msg.sender.call.value(amount)());
   /* At this point, the caller's code is executed, and can call transfer() */
   balance[msg.sender] = 0;
  }
\end{lstlisting}
\begin{figure}
    \centering{\includegraphics[width=0.49\textwidth]{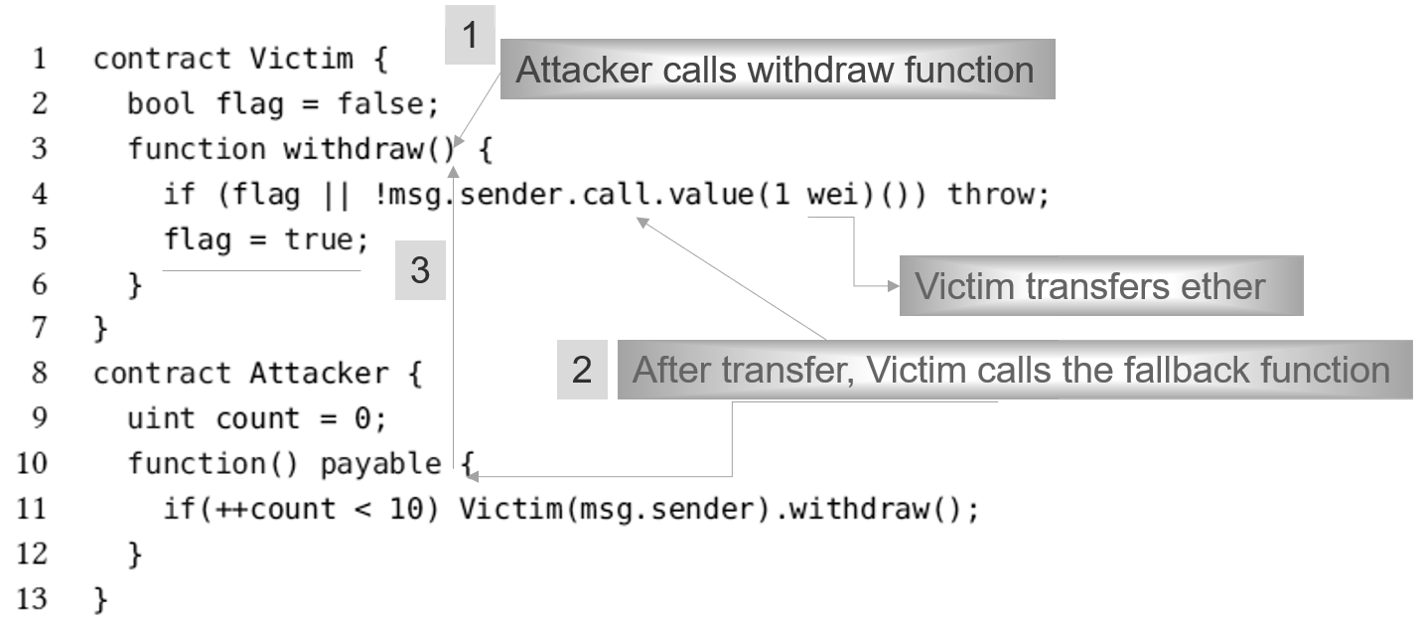}}
    \caption{The DAO Attack}
    \label{fig:daoattack}
\end{figure}
\section{Motivation}
The Decentralized Autonomous Organization (known as the DAO) was initiated in May 2016 as a venture capital fund for the crypto and decentralized space\cite{surveyattacks}, built as a smart contract on Ethereum blockchain concept. The lack of a centralized authority aided in running this organization at a reduced cost and it provided more control and access to the investors that participated. During the creation period of the DAO, anyone could send Ether to a unique wallet address in exchange for DAO tokens. This creation period gathered 12.7M Ether which was worth around USD 150M at the time, making it a huge success in the Ethereum world. Essentially, this platform was created to fund anyone with a project idea deemed potentially profitable by the community. Anyone with DAO tokens could vote on the pitch and receive rewards in return if the projects turned a profit.
However, on June 17, 2016, a hacker was able to attack this Smart contract by exploiting a vulnerability in the code that allowed him to transfer funds from the DAO. As reported by M. Saad et al. \cite{DBLP:journals/corr/abs-1904-03487} approximately, 3.6 million Ether was stolen, the equivalent of USD 70M at the time.
The attacker was able to request the smart contract (DAO) to give the Ether back multiple times before the smart contract could update its balance. This was possible because of the fact that when the DAO smart contract was written, the developers did not consider the possibility of a recursive call and the fact that the smart contract first sent the ether and then updated the internal token balance.
The reentrancy vulnerability exploitation in the DAO attack(as shown in Figure \ref{fig:daoattack}) was accomplished in four steps,
\begin{enumerate}
    \item The Attacker initiates a transaction by calling withdraw function of Victim;
    \item The Victim transfers the money and calls the fallback function of the Attacker;
    \item The fallback function recursively calls the withdraw function again, i.e., \textit{Reentrancy};
    \item Within an iteration bound, extra ether will be transferred multiple times to the Attacker.
\end{enumerate}
\begin{figure}[!t]
\centerline{\includegraphics[width=.42\textwidth]{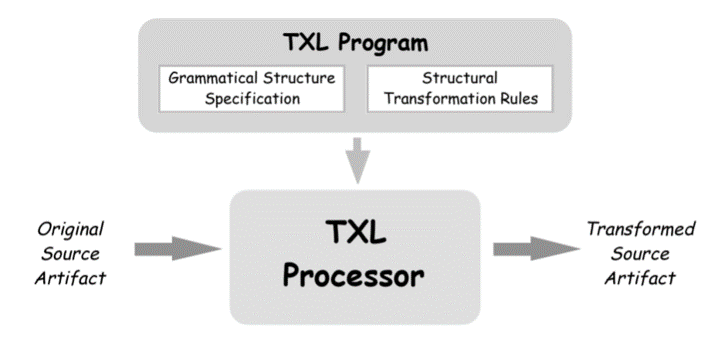}}
\caption{Txl Paradigm\cite{TXL}}
\vspace{-0.5 cm}
\label{fig:txl}
\end{figure}

\section{Proposed Framework}

As the example in Figure \ref{fig:daoattack} demonstrates, there are two parts to check for vulnerability. First, a call to an external function is executed. Second, there is a write to a persistent state variable after the external call.
Therefore, the pattern in the proposed analysis solution should combine both static and dynamic analysis methodologies to look for a persistent account state update after an external call to an untrusted address statically and to check if the recursive calling of such a function is possible dynamically.
\begin{figure*}[!t]
\centerline{\includegraphics[width=.8\textwidth]{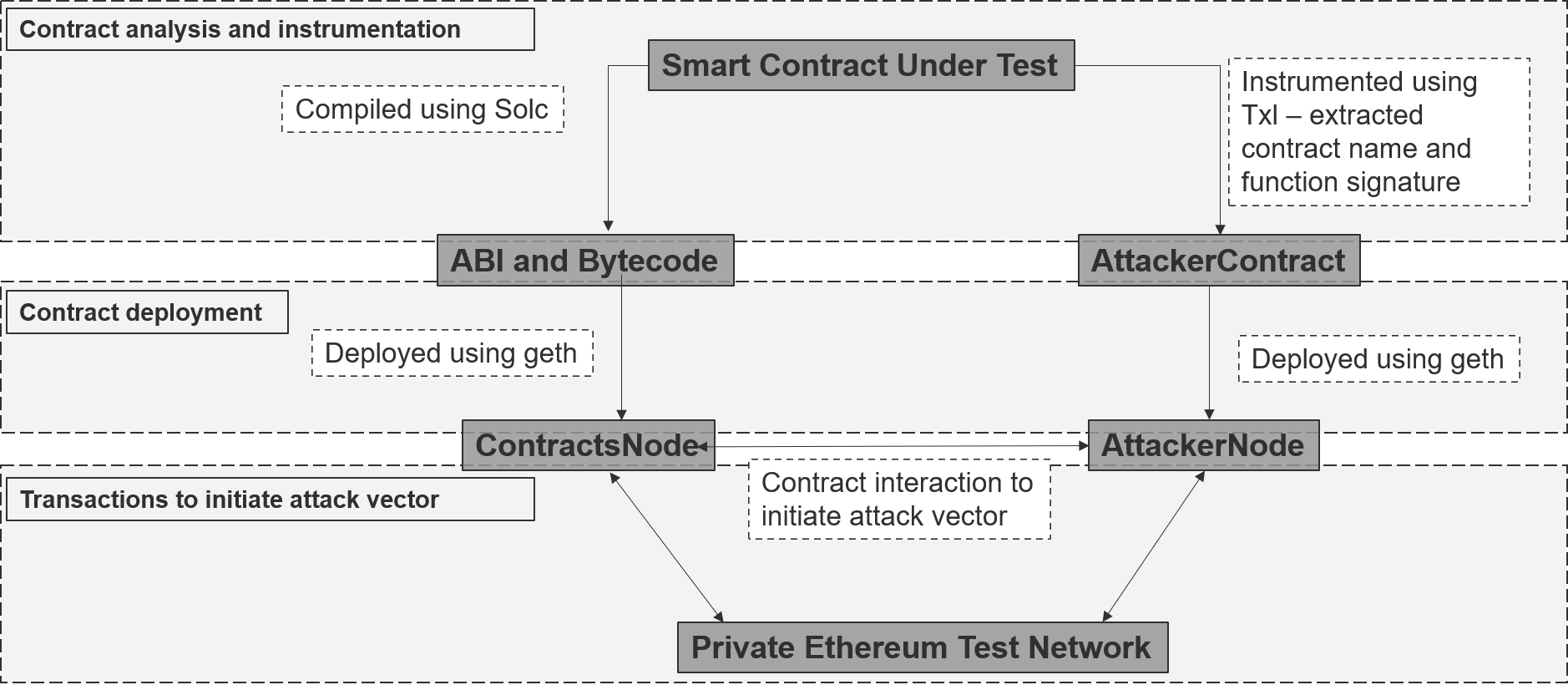}}
\caption{Proposed Framework}
\vspace{-0.2 cm}
\label{fig:architecture}
\end{figure*}
Our framework combines static and dynamic analysis to better capture the above-described attack scenario. The first stage of our framework uses static analysis and automated instrumentation of the smart contract under test and to efficiently and accurately capture the possible attack scenarios. 

The second stage uses the information obtained from stage one to automatically generate an attacker contract which will be used in a later stage to interact with a deployed version of the smart contact under-test and that do demonstrate dynamically whether the smart contract under-test has a confirmed \textit{Reentrancy} vulnerability.   

\subsection{Contract Analysis and Instrumentation}
To overcome the challenges in detecting \textit{Reentrancy} vulnerability in Smart-Contracts, the proposed framework, presented in Figure \ref{fig:architecture}, utilizes Txl\cite{TXL},  a programming language used for structural analysis and source transformation. The TXL paradigm (see Figure \ref{fig:txl}) consists of a grammar and transformation rules that are used to parse the input text into Abstract Syntax Tree (AST), and this intermediate AST is transformed into a new AST of the target domain, and finally\textcolor{red}{,} this output is unparsed to a new output text. 

In our framework, we used TXL to automate the following tasks: 
\begin{enumerate}
\item To parse the smart contract under test to identify potentially vulnerable functions, and to extract their names to help generate attacker contracts. Those contracts were aimed to be used in a later stage to create a \textit{Reentrancy} attack scenario. 
\item To instrument the contract and that to analyze the run-time behaviour of the smart contract under test during the attack.
\end{enumerate}
However, the existing TXL grammar resources to be used with TXL did not have one for Solidity programming language, this called for the creation of a TXL grammar for Solidity\cite{Solidity} to enable parsing of solidity smart contracts using TXL. 

The potentially vulnerable functions were determined if they contained an external call using any of these three Solidity’s built-in functions: \textit{transfer()}, \textit{call()}, \textit{send()}. These function signatures were extracted using TXL and outputted in a .txt file. This output file was used to modify the Attacker Contract to exploit \textit{Reentrancy} vulnerability in the smart contract under test if it existed.

An \textit{AttackerContract}, as shown in Listing\ref{lst8}, is automatically created to interact with potentially vulnerable functions of the smart contract under test with a \textit{Reentrancy} attack scenario. For illustration, \textit{FairDare} smart contract is used as an example from the collected dataset for this research. This framework successfully detected \textit{Reentrancy} vulnerability within it in our experiment.

As shown in Listing\ref{lst8}, the \textit{AttackerContract} tries to request ether multiple times from the smart contract under test with \textit{Reentrant} attack. At first, the AttackerContract creates an instance of the \textit{FairDare} contract and thus calls the constructor of the \textit{FairDare} contract. It then calls the withdraw()  function of \textit{FairDare} smart contract to initiate the attack (see Listing \ref{lst9}). Within the withdraw() function, the \textit{FairDare} smart contract sends ether with a transfer() before setting the value of the corresponding \textit{depositAmount} to 0. Since the call has no parameters provided, it will invoke the callback function of the \textit{AttackerContrac}t. 

\begin{lstlisting}[caption={AttackerContract},label={lst8}]
import "FairDare.sol";
contract AttackerContract_FairDare{
    address payable private _owner;
    address payable private _vulnerableAddr;
    FairDare public fd =FairDare(_vulnerableAddr);
    constructor() public {
         _owner = msg.sender;
    }
    function() external{
        fd. Withdraw();
    }
    function transferToOwner() public {
        _owner.transfer(address(this).balance);
    }
}
\end{lstlisting}
Within the callback function, the \textit{AttackerContract} can invoke the withdraw() function again as the \textit{Reentrancy} call. As a result, the \textit{FairDare} smart contract will send ether to the \textit{AttackerContract} again until all its ether is depleted.
With the help of the \textit{AttackerContract}, we can verify if the potential \textit{Reentrancy} vulnerability detected by TXL pattern matching is exploitable.

\begin{lstlisting}[caption={Withdraw() function of FairDare Smart Contract (Smart contract under test)\cite{Etherscsan}},label={lst9}]
function withdraw() public {
    require(tx.origin == msg.sender, "");
    uint blocksPast = block.number - depositBlock[msg.sender];
    if (blocksPast <= 100) {
        uint amountToWithdraw = depositAmount[msg.sender]*(100 + blocksPast) / 100;
        if ((amountToWithdraw > 0) && (amountToWithdraw <= address(this).balance)){
            msg.sender.transfer(amountToWithdraw);depositAmount[msg.sender] = 0;
        }
    }
}
\end{lstlisting}

\subsection{Contracts Deployment}
For the contract deployment stage, we need to choose an Ethereum Private Test Network Specifications. In this framework, we use Geth\cite{Geth}, a Golang implementation of Ethereum, to create a private blockchain. This enabled us to create a new, and private blockchain from scratch which we will be using to deploy and test any smart contract. This private blockchain will not be connected to the Ethereum main net and therefore does not require real ether.
To use Geth\cite{Geth}, we created a new account that represents a key pair. 
We simulated having multiple computers storing our blockchain by creating two nodes, which act like two different computers hosting and interacting with the same blockchain. Each node that wants to interact with a blockchain will store the blockchain on their computer. We initiated these two nodes from Geth using two different Terminal windows to simulate two different computers. We used two different data directories, so each node has a separate place to store their local copy of our blockchain as shown in the below Genesis Block,

\begin{lstlisting}[caption={CustomGenesis.json file},label={lst10}]
{
 "config": 
    {
        "chainID": 1708,
        "homesteadBlock": 0,
        "eip150Block": 0,
        "eip155Block": 0,
        "eip158Block": 0
    },
 "alloc": 
    {
      "0xB1C0a62c5df3AE6469031D5BC0842382187C7F25": 
        {
        "balance": "100000000000000000000000000000"
        }
    },
 "difficulty": "0x4000",
 "gasLimit": "0xffffffff",
 "nonce": "0x0000000000000000",
}
\end{lstlisting}
Our CustomGenesis.json (See Listing \ref{lst10}) file determines:
\begin{enumerate}
    \item Contents of the genesis block, or the first block of our blockchain.
    \item Configuration rules that our blockchain will adhere to. 
    \item homesteadBlock: defines the version of the Ethereum platform, we set this attribute to 0 as we are already on the version of Homestead version.
    \item  eip150Block/eip155Block/eip158Block: Ethereum Improvement Proposals (EIPs) indicate hard-forking for backward-incompatible protocol changes, we set this attribute to 0 as our private network will not require this hard-forking.
    \item difficulty: On our test network, we kept this value low to avoid waiting during tests, since the generation of a valid Block is required to execute a transaction on the Blockchain.
    \item gasLimit: Gas is Ethereum's fuel that is spent during transactions. We set this value high enough in our test network to avoid being limited during tests.
    \item alloc: We used this attribute to create our wallet and prefill it with fake ether.
\end{enumerate}

\subsection{Interacting with Blockchain}
To interact with the blockchain, we launched the geth \textit{console} command to initiate the ports for ContractsNode and AttackerNode. 
 After simulating the contracts node and attacker node on our private Ethereum network, the contracts under test are deployed on these nodes in the following steps:
\begin{enumerate}
    \item Compiling the contract under test using either “solcjs” command or in “remix online IDE"\cite{Remix}. After compiling the contract, the ABI and bytecode of the contract under test are generated.
 \begin{lstlisting}[caption={ABI code of FairDare smart contract},label={lst11}]
[
    {"constant": false, 
    "inputs": [],
    "name": "withdraw",
    "outputs": [],
    "payable": false,
    "stateMutability": "nonpayable",
    "type": "function" [10] 
    },
    {"payable": true,
     "stateMutability": "payable",
     "type": "fallback"
    }
]
    \end{lstlisting}
    \item Using the ABI and bytecode, the contract is deployed on the network.
    \item After deploying the contract onto a node, the transaction is submitted to all the peers of the blockchain. The miner.start() command starts the mining process and miner.stop() ends this process. After the mining process is completed, the contract gets added to the blockchain. And, once added to the blockchain, this contract cannot be changed.
\end{enumerate}
Since we set the \textit{--nodiscover} flag while connecting to our blockchain, these nodes cannot automatically interact with other nodes. Therefore, to enable transactions, we had manually tell one of the nodes about the other node as follows using the \textit{admin.nodeInfo.enode()} and \textit{admin.addPeer()} commands.
After connecting both the nodes, transactions can be sent from an account to another using:

\vspace{0.3 cm}
\textit{\small{eth.sendTransaction ({ from: eth.accounts[0], to: “”, value:"" })}}
\vspace{0.3 cm}

A specific function of a contract can be called using,
\vspace{0.3 cm}

\textit{\small{contractAddress.call.value ("ether to be transferred") .gas ("amount of gas to be spent") (abi.encodeWithSignature ("functionName(types)" , parameter values));}}
\vspace{0.3 cm}


Where, \textit{ ContractAddress} is the address generated when a contract is deployed,
\textit{abi.encodeWithSignature} is used to encode structured data to aid in building valid call to functions.

The function name was extracted using TXL and substituted in the above command and the \textit{AttackerContract} as well. Table\ref{functionDet} lists the extracted function name for each contract under test.
\begin{table}[!t]
\caption{Function Name extracted from contracts under test}
\begin{center}
 \begin{tabular}{||c c||} 
 \hline
 Contract Name & Function Name\\ [0.5ex] 
 \hline\hline
 DeFi &	withdraw() \\
 \hline
 Globalcryptox & constructor() \\
 \hline
 FairDare & withdraw() \\
 \hline 
 Moneybox & withdraw()  \\
 \hline
 AIRToken  & burn() \\
 \hline 
 QuizBLZ & try() \\ [1ex] 
 \hline
\end{tabular}
\end{center}
\label{functionDet}
\end{table}

\begin{figure*}[!t]
\centerline{\includegraphics[width=.99\textwidth]{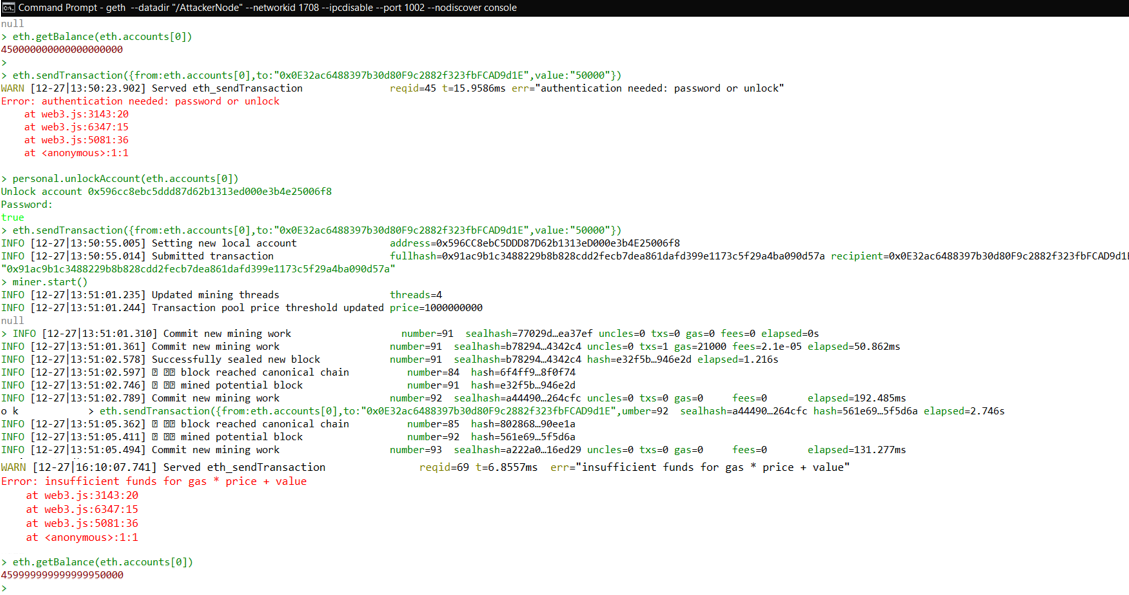}}
\caption{Ether balance after submitting transaction and after mining}
\vspace{-0.2 cm}
\label{fig:AfterMining}
\end{figure*}

The complete process of deploying contracts and sending transactions can be depicted as follows:
After successfully initiating transaction between \textit{FairDare} smart contract and the modified \textit{AttackerContract} for \textit{FairDare} smart contract, the \textit{Reentrancy} vulnerability in the smart contract under test was exploited as the fallback function of AttackerContract called the withdraw () function of the \textit{FairDare} smart contract recursively till insufficient funds exception was thrown and only the last call to withdraw () function was reverted. The ether balance was tracked to monitor the \textit{Reentrancy} bug existence as the ether of \textit{FairDare} smart contract gets depleted multiple times and ether balance at the attacker node increases more than what it had requested. (See Figure \ref{fig:AfterMining})
\begin{figure}[!t]
\centerline{\includegraphics[width=.49\textwidth]{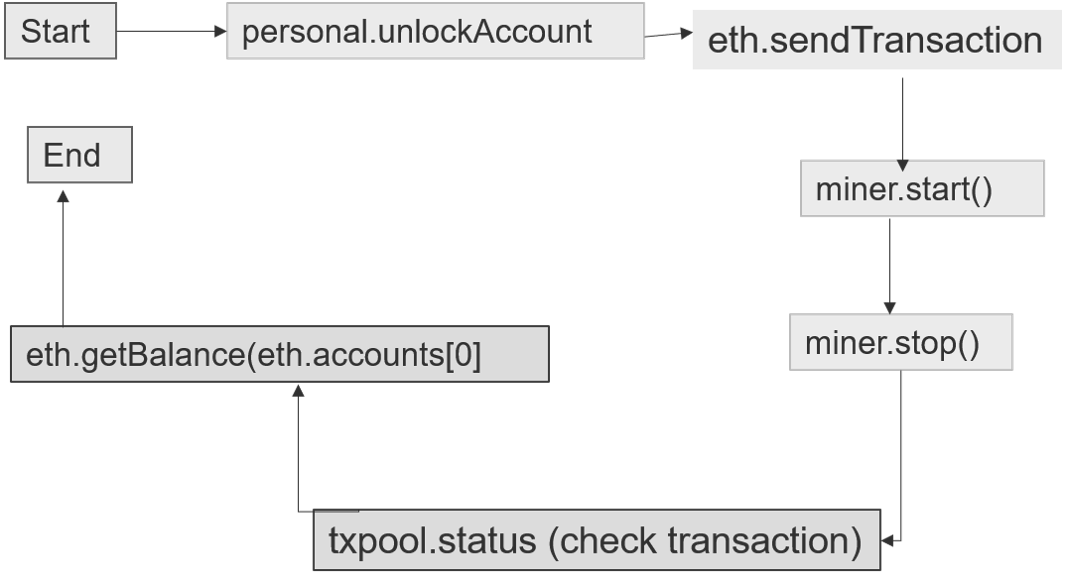}}
\caption{Blockchain process of sending a transaction}
\vspace{-0.2 cm}
\label{fig:BlockchainProcess}
\end{figure}

\begin{table} [!b]
\caption{Details of  6 modified contracts under test}
\begin{center}
 \begin{tabular}{||c c c c||} 
 \hline
 Contract Name & Compiler Version & Balance & LOC \\ [0.5ex] 
 \hline\hline
 DeFi &	V0.5.12 & 0.036 ether &	449 \\
 \hline
 Globalcryptox &V0.4.25& 0 ether & 235\\
 \hline
 FairDare & V0.5.12	&0.0071 ether & 41\\
 \hline 
 Moneybox & v0.5.13	& 0 ether &	40 \\
 \hline
 AIRToken  & V0.5.13	& 0 ether	& 304 \\
 \hline 
 QuizBLZ & v0.5.12  & 0 ether & 54 \\ [1ex] 
 \hline
\end{tabular}
\end{center}
\label{contractDet}
\end{table}
\section{Related Work}
 {di Angelo} and {Salzer} \cite{8813282} presented a state of the art survey on tools for Ethereum Smart Contracts analysis. For a detailed comparison of those tools, the reader is advised to review this survey\cite{8813282}. 
 In our paper, we only discuss two tools that are closest to our approach, Contract Fuzzer \cite{ContractFuzzer}, and Reguard\cite{Reguard}.
\subsection{Contract Fuzzer}
Contract Fuzzer is a tool developed by Jiang et al. \cite{ContractFuzzer} to test the Smart-Contracts for vulnerabilities using a fuzzing technique. To detect the vulnerabilities, this tool starts with an initial analysis of the interfaces that the Smart-Contract exposes, it then randomly develops fuzzing inputs for these interfaces and observes the execution logs of the application.
This tool targets more than one known vulnerabilities in Smart contracts, however, focusing on detecting \textit{Reentrancy} vulnerabilities in smart contracts it may have slower performance compared to our proposed framework because it triggers test transactions against all the functions of a smart contract and not the potentially vulnerable functions by parsing the contract under test early on.
\subsection{Reguard}
ReGuard is a dynamic analysis tool to detect \textit{Reentrancy} vulnerabilities in Smart-Contracts developed by Liu et al. \cite{Reguard}. This tool tests the Smart-Contracts by initially transforming the Smart-Contract code into C++ and then generating fuzzing inputs to recreate Blockchain transactions as possible attacks. Then, ReGuard performs vulnerability detection through dynamic analysis.
In comparison with this tool, we believe our framework will provide better performance as static analysis of smart contracts is done in its original form, i.e. using a context-free grammar for Solidity and not by transforming the smart contracts into some other programming language and subjecting the transformed contract to vulnerability test.
\section{Evaluation}
The dataset we used for the evaluation was extracted from Etherscan\cite{Etherscsan}, a free-to-use platform for Blockchain analytics based on Ethereum. It is essentially a Block Explorer that allows users to easily lookup, confirm and validate transactions that have taken place on the Ethereum Blockchain. Smart-Contracts developers can get a substantial benefit from APIs available that can be utilized to either build decentralized applications or serve as a dataset for Ethereum Blockchain analysis.

We modified 6 contracts from Etherscan to test for \textit{Reentrancy} vulnerability. The modification was done to introduce \textit{Reentrancy} vulnerability in the selected contracts by updating the token balance after transferring the ether to an external contract. Table \ref{contractDet} shows the specifications of the contracts used.

\begin{table}[!b]
\caption{Results of \textit{Reentrancy} vulnerability analysis. FP: false positive. FN: false negative. * : Analysis failed}
\begin{center}
 \begin{tabular}{||c c c||} 
 \hline
 Contract Name & ContractFuzzer & OurFramework\\ [0.5ex] 
 \hline\hline
 DeFi &	(1-FP)2 & 1 \\
 \hline
 Globalcryptox & 1 & * \\
 \hline 
 Moneybox & 2 & 2  \\
 \hline
 AIRToken  & (2-FP) 3 & 1 \\
 \hline 
 FairDare & 1 & 1 \\ [1ex] 
 \hline
 QuizBLZ & * & 1 \\ [1ex] 
 \hline
\end{tabular}
\end{center}
\label{ComparisonDet}
\end{table}

We have conducted a preliminary evaluation of our proposed framework on 6 modified smart contracts from Etherscan\cite{Etherscsan} and our framework was able to detect the \textit{Reentrancy} vulnerability in all of our modified contracts, with an exception of one contract that was caused due to version incompatibility of the context-free grammar created for Txl parser. We are evolving our TXL grammar to account for the various versions of Solidity.

We used Reguard\cite{Reguard} and ContractFuzzer \cite{ContractFuzzer} to compare and to evaluate the accuracy of the proposed framework. While ContractFuzzer allows checking on a set of predefined vulnerabilities patterns in smart contracts, we focused on \textit{Reentrancy} vulnerability in this evaluation. As ContractFuzzer targets every function of a contract under test, it covers a wider attack surface but produced more false positives compared to our framework which included analyzing a contract statically and targeting potentially vulnerable functions only which lead to reduced false positives (See Table \ref{ComparisonDet}). We were not able to access the Reguard tool to analyze our dataset but we believe our proposed framework may return better results in performance than Reguard because, in our framework, Smart contracts are directly analysed and not transformed into a C++ intermediate representation.
\section{Conclusion}
In this paper, a combined static and dynamic analyzer framework is proposed to detect \textit{Reentrancy} vulnerabilities in Ethereum smart contracts. This framework leverages TXL for analyzing these smart contracts to extract the function signature of potentially vulnerable functions in a contract.
Then, the \textit{Reentrancy} bug detection is done at the run-time while interacting with the vulnerable contract via an attacker contract and that by trying to recreate the \textit{Reentrancy} scenario.
\section{Future Work}
The future work in this research would be to deploy our approach as a tool and that by automatically integrating the three automated stages of our approach (Analysis, Deploy and Test) of smart contracts for \textit{Reentrancy} vulnerability. This automated tool would also be scalable to include Solidity smart contracts of newer versions. Currently, this framework targets Solidity smart contracts of versions after v4.28. The Dataset used in this work utilizes 6 modified contracts from Etherscan [8]. However, this can be increased to test many more smart contracts by developing a web scraper and running on the Etherscan portal to collect many smart contracts deployed on this website.
\vspace{-0.1 cm}
\section*{Acknowledgments}
This work is supported in part by the Natural Sciences and Engineering Research Council of Canada (NSERC).

\balance
\bibliographystyle{IEEEtran}
 \bibliography{Noama2019.bib}
\nocite*
\end{document}